\documentclass{llncs}
\usepackage{url}
\usepackage{algpseudocode}
\usepackage{listings}
\usepackage{graphicx}
\usepackage{amsmath}
\usepackage{floatrow}
\DeclareFloatFont{scriptsize}{\scriptsize}
\usepackage[caption=false]{subfig}
\usepackage{amssymb}
\usepackage{paralist}
\usepackage{xspace}

\newcommand{\etc}{\textit{etc.}}
\newcommand{\ie}{\textit{i.e.}}
\newcommand{\etal}{\textit{et al.}~}
\newcommand{\eg}{\textit{e.g.}}
\newcommand{\cf}{\textit{cf.}}

\begin{document}

\title{Bayesian Attack Model for Dynamic Risk Assessment}
\author{Fran\c{c}ois-Xavier Aguessy\inst{1}$^{,}$\inst{2} \and Olivier Bettan \inst{1} \and Gregory Blanc\inst{2} \and Vania Conan\inst{1} \and Herv\'e Debar\inst{2}\\ \email{francois-xavier.aguessy@telecom-sudparis.eu}}

\institute{Thales Group, 4 avenue des Louvresses, 92622 Gennevilliers, France
\and SAMOVAR, T\'el\'ecom SudParis, Universit\'e Paris Saclay, 9 rue Charles Fourier, 91011 Evry, France}

\maketitle

\begin{abstract}
	Because of the threat of advanced multi-step attacks, it is often difficult for security operators to completely cover all vulnerabilities when deploying remediations.
	Deploying sensors to monitor attacks exploiting residual vulnerabilities is not sufficient and new tools are needed to assess the risk associated to the security events produced by these sensors.
	Although attack graphs were proposed to represent known multi-step attacks occurring in an information system, they are not directly suited for dynamic risk assessment.
	In this paper, we present the Bayesian Attack Model (BAM), a Bayesian network-based extension to topological attack graphs, capable of handling topological cycles, making it fit for any information system.
	Evaluation is performed on realistic topologies to study the sensitivity of its probabilistic parameters.
\end{abstract}


\section{Introduction}

Managing the security of Information Systems (IS) is increasingly complex, due to the numerous security mechanisms that are implemented, and the significant amount of dynamic data produced by security enforcement points.
In critical environments, security operators generally know most of the vulnerabilities of their IS thanks to regular vulnerability scans.
Unfortunately, many vulnerabilities are not patched, either because patching may disrupt critical services, or because they are not a priority for system administrators.
As a second line of defence, security operators deploy sensors (\eg, Host or Network Intrusion Detection Systems) generating alerts when an attacker attempts to exploit such vulnerabilities.
As these security events are produced, operators need to evaluate the risk brought by ongoing attacks in their system, to respond appropriately: this process is called dynamic risk assessment (DRA) \cite{lopez2013dynamic}.

The most impacting attacks are composed of several successive exploitation steps.
Several models have been proposed to formalize such multi-steps attacks.
An attack graph is a model regrouping all the paths an attacker may follow in an information system.
It has been first introduced by Phillips and Swiler in~\cite{Phillips:1998cy}.
A study of the state of the art about attack graphs compiled from early literature on the subject has been carried out by Lippmann and Ingols~\cite{lippmann2005annotated}, while a more recent one was made available by Kordy \etal~\cite{Kordy:2013ty}.
Topological attack graphs are based on directed graphs.
Their nodes are topological assets (hosts, IP addresses, etc.) and their edges represent possible attack steps between such nodes~\cite{Jajodia2005}.
Attack graphs are generated with attack graph engines.
There are three main attack graph engines:
\begin{inparaenum}[(1)]
\item {\sl MulVAL}, the Multi-host, Multi-stage Vulnerability Analysis Language tool created by Ou \etal~\cite{Ou2005},
\item the Topological Vulnerability Analysis tool ({\sl TVA}) presented by Jajodia \etal~in~\cite{Jajodia2005,Jajodia2011} (commercialized under the name \emph{Cauldron}) and
\item Artz's {\sl NetSPA}~\cite{Artz2002}.
\end{inparaenum}
Attack graphs are attractive because they leverage readily available information (vulnerability scans and network topology).
However, they are not adapted for ongoing attacks, because they can not represent the progression of an attacker nor be triggered by alerts.
Thus, they must be enriched to provide the functionalities needed to perform Dynamic Risk Assessment, for example using Bayesian networks.

A Bayesian network is a probabilistic graphical model introduced by Judea Pearl~\cite{pearl1986fusion}.
It is based on a Directed Acyclic Graph, where nodes represent random variables, and edges represent probabilistic dependencies between variables~\cite{ben2007bayesian}.
For discrete random variables, these dependencies can be specified using a Conditional Probability Table associated with each child node.
Bayesian networks are particularly interesting for computing inference, \ie~calculating the probability of each state of all nodes of the network, given some evidences, \ie~nodes that have been set to a specific state.
Inference can be done efficiently using the algorithm of Lauritzen and Spiegelhalter~\cite{lauritzen1988local}.
A Bayesian attack graph, introduced by Liu and Man in~\cite{Liu:2005un} is an extension of an attack graph based on a Bayesian network, constituted of nodes representing a host in a specific system state (a true state means that the host is compromised) and edges representing possible exploits that can be instantiated from a source host to a target host.
The major concern of building such a Bayesian network from an attack graph is due to the structure of a Bayesian network that must be acyclic, while attack graphs almost always contain cycles.
To avoid cycles, Liu and Man assume that an attacker will never backtrack once reaching a compromised state, but do not detail how such assumption is used to build the model.
In~\cite{Frigault:2008jp}, Frigault and Wang use Bayesian inference in Bayesian Attack Graphs to calculate security metrics in an information system.
Xie \etal~present in~\cite{Xie:2010ks} a Bayesian network used to model the uncertainty of occurring attacks.
The Bayesian attack graph is enhanced with three new properties: separation of the types of uncertainty, automatic computation of its parameters and insensitivity to perturbation in the parameters choice.
This model also adds nodes dedicated to dynamic security modelling: an \emph{attack action node} models whether or not an attacker action has been performed, a \emph{local observation node} models the inaccuracy of observations.

In this paper, we propose a new model combining attack graphs and Bayesian networks for DRA.
It is built from the knowledge security operators have about their IS: network topology, known vulnerabilities and detection sensors.
Then, we change the states of the sensor nodes according to the security events received.
This model is capable of representing the attacks that may occur (vulnerabilities) and the ones ongoing (alerts).
It outputs probabilities that attacks have succeeded and that assets of the IS may have been compromised.
With respect to the current state of the art, our contributions are twofold.
First, we provide an explicit model and process for handling cycles.
This process is supported by a clear definition of a set of model parameters.
The sensitivity of the model toward these parameters is studied in the validation.
Second, we provide a significant performance improvement in terms of number of nodes and vulnerabilities over the existing state of the art.
While classic Bayesian attack graph models are usually demonstrated over a few nodes and vulnerabilities, we show that our model can be realistically computed at the scale of an enterprise IS.

This paper is organised as follows: in Section~2, we formally define the structure and the conditional probability tables of our Bayesian Attack Model built from a topological attack graph.
Section~3 validates the results of the Bayesian Attack Model on a realistic use case and analyses its sensitivity toward the probabilistic parameters.
Section~4 compares our work with the related work, before concluding and presenting future work, in Section~5.


\section{The Bayesian Attack Model}

Given the advantages brought by Bayesian Attack Graphs, they provide a strong foundation for dynamic security modelling.
Our proposal extends Bayesian Networks to be used for DRA with real-scale IS.

The Bayesian Attack Model (BAM) described all along this section is built from a Topological Attack Graph, which is described in section \ref{sec:tag}, and a set of detection alerts.
The BAM is composed of submodels called Bayesian Attack Trees (BAT).
BAT and BAM are described in section \ref{sec:bam}.
Each BAT is composed of a sequence of attack steps, typed nodes linked together.
They are described in section \ref{attack-step-structure}.
The probabilistic relations between nodes of a BAT are described in conditional probability tables whose content is detailed in section \ref{sec:cpt}.

\subsection{Topological Attack Graph}
\label{sec:tag}
The BAM is built from a topological attack graph.

\begin{definition}
  \label{def-tag}
	A \textbf{topological attack graph} is a directed graph $\emph{TAG(TN,AS)}$:
	\begin{itemize}
		\item $TN=
		\{\emph{TN}_i
		, i\in\{1..N\}\}$ is a set of $N$ \textbf{topological nodes}: the assets of an information system,
		\item $\emph{AS}$ is a set of \textbf{attack steps}, the edges that represent the fact that an attack allows the attacker to move from the parent topological node to the child topological node.
	    \begin{itemize}
				\item Each attack step has a \textbf{type} of attack, describing how the attacker can move between nodes (exploitation of a vulnerability, credential theft,~\etc).
	      \item Depending on the type of attack, each attack step is associated with a set of \textbf{conditions} $\emph{[c]}$.
	      \item Some attack steps are associated with a \textbf{sensor} that may raise an alert indicating that this attack has been detected.
	    \end{itemize}
	\end{itemize}
\end{definition}

A TAG can be generated with an attack graph engine such as MulVAL~\cite{Ou2005} or TVA~\cite{Jajodia2011}.
Topological nodes represent, for example, an IP address or a computer cluster.
Attack steps are, for example, the exploitation of a vulnerability.

\begin{definition}
  \label{def-condition}
	A \textbf{condition} $\emph{c}$ is a fact that needs to be verified, for an attack step to be possible. It is associated with a probability of success $\emph{P(c)}$.
\end{definition}

The condition fact is, for example, ``a vulnerability is exploited on the destination host''.
For such conditions, in our experiments, we use an approximation of the probability of successful exploitation using information coming from the Exploitability Metrics of the Common Vulnerability Scoring System (CVSS)~\cite{first2015common}.
It is deduced from
\begin{inparaenum}[(1)]
\item the Attack Complexity (AC),
\item Privileges Required (PR),
\item and User Interaction (UI) values,
\end{inparaenum}
as well as the Attack Vector (AV), which is taken into account when constructing the topological attack graph.

\begin{definition}
  \label{def-sensor}
	A \textbf{sensor} $\emph{s}$ of an attack step is an oracle issuing an alert when the attack step has been detected. It is associated with a false negative and a false positive rates.
\end{definition}

A sensor represents, for example, an Intrusion Detection System, a System Event Management, or a human report.

\subsubsection{Grouping attack steps}
\label{sec:grouping-as}
In topological attack graphs, there may exist many attack steps between two topological nodes.
Attack steps can be of different types, depending on the attack (\cf~Definition~\ref{def-tag}).
Generally, there are very few possible types of attack steps (\eg, the remote exploitation of a vulnerability on a server).
In order to reduce the size of the model, while preserving information, we group all attack steps of the same type between two topological nodes into a single vertex with
\begin{inparaenum}[(1)]
	\item a new condition: a multivariable boolean function (usually, an OR) of all conditions applying to the grouped attack steps;
	\item an attached sensor node activated only when the boolean function of grouped sensors is true.
\end{inparaenum}

When several conditions $c_i$ of an attack step $as$ are grouped in one condition $c$, we define the probability of successful exploitation associated with this new condition.
For example, when grouping several conditions $c_i$ ``a vulnerability is exploited on the destination host'' into one new condition $c$ ``at least one vulnerability of the list is exploited on the destination host'', we assume that the exploitation of each vulnerability is independent, to compute its probability of exploitation $P(c)$.
This is an acceptable approximation since we consider all the existing vulnerabilities between two topological nodes.
Thus, the probability of exploitation $P(c)$ becomes:

\begin{equation*}
  P(c)=P(\bigvee \limits_{i\in \{\mbox{\scriptsize{vulnerabilities of} }as\}}c_{i})
	= 1 - \prod_{i \in \{\mbox{\scriptsize{vulnerabilities of} } as\}}{(1 - P(c_i))}
\end{equation*}

\subsubsection{Breaking cycles in topological attack graphs}

A TAG is a model defined globally for a system, containing all potential attacks that can happen.
It thus almost always contains cycles, especially inside local networks in which any host can attack any other one.
For example, a host $tn_1$ may be able to attack another host $tn_2$ that can also attack $tn_1$ (directly or in several steps).
A common assumption to break cycles in attack graphs is that an attacker will not backtrack, \ie, come back on a node he has already successfully exploited.
This is reasonable because backtracking does not bring new information about attack paths.
It has been properly justified by Ammann~\etal in~\cite{ammann2002scalable} and by Liu and Man in~\cite{Liu:2005un}.
However, the solutions of the state of the art for Bayesian modelling of an attack graph such as the ones of Liu and Man \cite{Liu:2005un} and Poolsappasit~\etal~\cite{poolsappasit2012dynamic} use this assumption to delete arbitrary possible attack steps.
In reality, it is impossible to know a priori which path the attacker can choose.
Deleting paths in the Bayesian model thus suppresses actually possible attacker actions.
The only way to break cycles, while keeping all possible paths, is to enumerate all paths, starting from every possible attack source, keeping in the nodes a memory of the path of the attacker.
So, using this memory, we build an acyclic TAG by ensuring that the paths do not backtrack on already exploited nodes.
For example, a node $tn_1tn_2tn_3$ means that the attacker controls the node $tn_3$, having first compromised $tn_1$, then $tn_2$, finally $tn_3$.
Unfortunately, this cycle breaking process causes a combinatorial explosion in the number of nodes of the model.
We discuss in Section~\ref{nbsteps-justification} how we mitigate such limitation.

\subsection{Representation of an attack step in BAM}
\label{attack-step-structure}

An attack step in the TAG is an edge which is associated with several conditions and can be related to a detection sensor.
In the BAM, we detail the attack steps, the conditions, and sensors as nodes, in order to model the probabilistic interactions between such elements, using the nodes detailed below.
Each node represents a boolean random variable with two mutually exclusive states.

\begin{definition}
  \label{def-btn}
  A \textbf{Bayesian topological node} $\emph{btn}$ $(\emph{tn}_1,\cdots, \emph{tn}_n)$, \\ with $\forall i,  \emph{tn}_i~\in~\emph{TN}$ (\cf~Def.~\ref{def-tag}), is a node of the BAM representing the random variable describing the state of compromise of $\emph{tn}_n$ using the path of the topological attack graph $\emph{tn}_1 \rightarrow \cdots \rightarrow \emph{tn}_n$ (\ie, $\emph{Compromised}$ or $\emph{NotCompromised}$).
\end{definition}

\begin{definition}
  A \textbf{Bayesian attack step node} $\emph{basn(as)}$, with $\emph{as} \in \emph{AS}$ \\(\cf~Def.~\ref{def-tag}), is a node of the BAM representing the random variable describing the attack success of $\emph{as}$ (\ie, $\emph{Succeeded}$ or $\emph{Failed}$).
\end{definition}

\begin{definition}
  A \textbf{Bayesian condition node} $\emph{bcn(c)}$, with $\emph{c}$ a condition \\(\cf~Def.~\ref{def-condition}), is a node of the BAM representing the random variable describing that the condition $\emph{c}$ is fulfilled (\ie, $\emph{Succeeded}$ or $\emph{Failed}$).
\end{definition}

\begin{definition}
	\label{def-bsen}
  A \textbf{Bayesian sensor node} $\emph{bsen(s)}$, with $\emph{s}$ a sensor (\cf~Def.~\ref{def-sensor}), is a node of the BAM representing the random variable describing the state of the sensor $\emph{s}$ (\ie, $\emph{Alert}$ or $\emph{NoAlert}$).
\end{definition}

These nodes are linked with edges, indicating that the child node has a conditional dependency to the state of its parents.
For example, a Bayesian attack step node has a dependency toward its condition(s) and the topological node from which it may be accomplished.
Thus it is the child of the nodes representing the conditions and the topological node.
In the same way, a Bayesian sensor node is the child of a Bayesian attack step, and a Bayesian topological node is the child of a Bayesian attack step.

\begin{definition}
	\label{def-edge}
  A \textbf{Bayesian edge} $\emph{e}$, is a link from a $parent$ node to a $child$ node that represents a conditional dependency of the $child$ toward its $parent$.
\end{definition}

Appendix~\ref{sec:appendix-detail-attack-step} Figure~\ref{bayesian-attack-step} shows the details of the representation of an attack step from $tn_n$ (source) to $tn_{n+1}$ (target).

\subsection{Complete Bayesian Attack Model}
\label{sec:bam}

\subsubsection{Bayesian Attack Tree and Global Model}

The complete BAM is composed of a family of Bayesian Attack Trees (BAT), as defined below, each one issued from one attack source.

\begin{definition}
	A \textbf{Bayesian Attack Tree} is a Bayesian network represented by $\emph{BAT(AS, DAG,P)}$ where:
	\begin{itemize}
		\item $AS$ is a special Bayesian Topological Node, the attack source of this BAT.
		\item $\emph{DAG(BN,E)}$ is a polytree structure, constituted of
		\begin{itemize}
			\item $BN$, the Bayesian nodes $\emph{BN=[btn],[basn],[bcn],[bsen]}$ (\cf~Defs.~\ref{def-btn}-\ref{def-bsen})
			\item $E$, the set of edges $E=\{e\}$ representing a conditional dependency between the nodes (\cf~Def.~\ref{def-edge}).
		\end{itemize}
	  \item $\emph{P}$ is a set of local probability distributions, associated with each node of $\emph{DAG}$. As all nodes are discrete random variables, the local probability distributions can be specified within a Conditional Probability Table.
	\end{itemize}
\end{definition}

To build the whole structure of one BAT of the BAM, we start from a potential attack source of the acyclic TAG.
It is the Attack Source and the root of the BAT.
Then, we recursively add the attack steps contained in the acyclic TAG with the nodes described in Subsection~\ref{attack-step-structure}.
To avoid cycles, each attack step is added, as soon as its target has not been already compromised during the currently followed path.
This can be achieved thanks to the memory of past topological nodes in Bayesian topological nodes.
This building process also ensures that the graph structure of each BAT is a polytree: a Directed Acyclic Graph for which there are no undirected cycles either.
This allows to use very efficient exact inference algorithms in the Bayesian network such as Pearl's algorithm~\cite{pearl1988probabilistic}.

The complete BAM is constituted of the set of all BATs.
As we consider that each topological node may be a source of attack, the BAM contains exactly $N$~BAT (\ie, the number of topological nodes in the TAG).

\begin{definition}
	The \textbf{Bayesian Attack Model} $\emph{BAM}(\{BAT_i\})$, is a family of $N$ Bayesian networks where, for all $i$ in $\{1..N\}, \emph{BAT}_i$ is a BAT, whose attack source is node $i$ in the topological attack graph.
\end{definition}

\begin{figure}[h]
	\begin{center}
		\includegraphics[width=12cm]{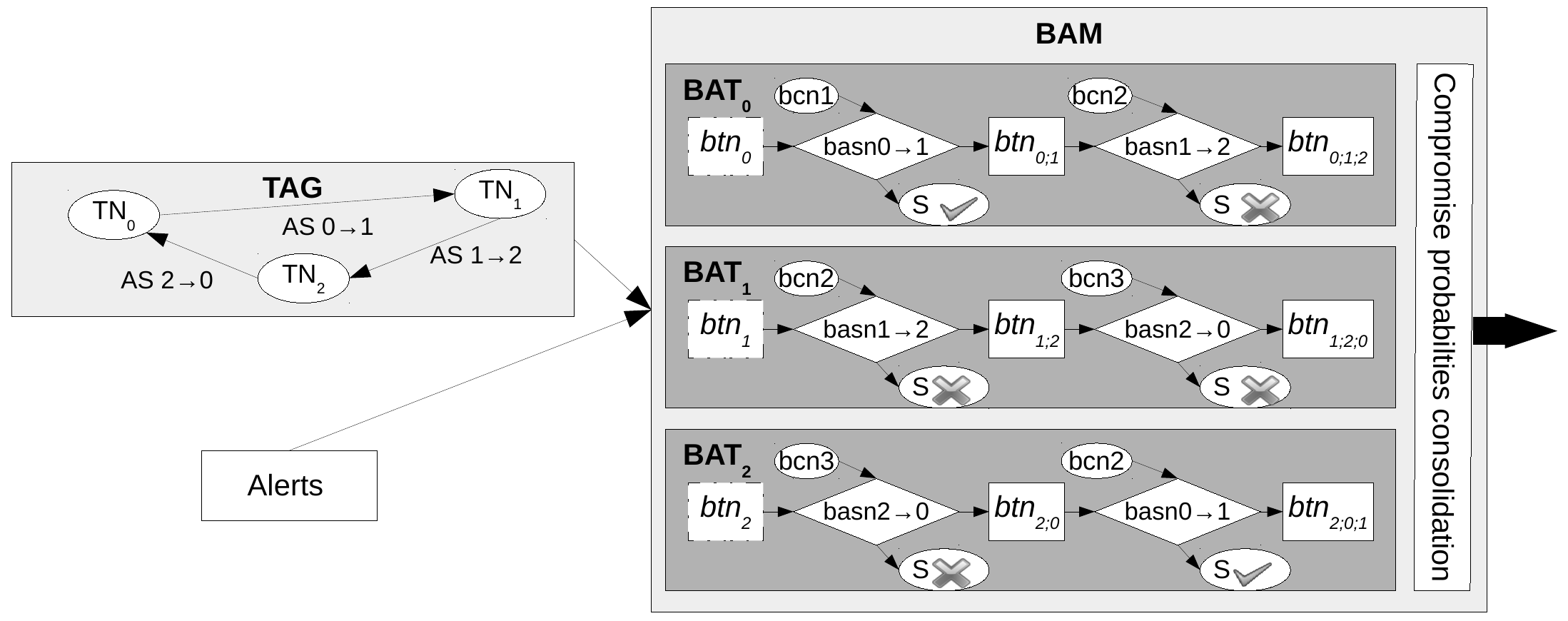}
	\end{center}
	\caption{\label{bam-architecture} Bayesian Attack Model Architecture}
\end{figure}

Figure~\ref{bam-architecture} summarises the global architecture of the BAM. In this example, it is built from a TAG containing 3 nodes and thus is composed of 3 BATs.

\subsubsection{Consolidation of probabilities}
\label{probability-consolidation}

As each Bayesian topological node contains the history of the attack that can lead to this node, many Bayesian topological nodes can represent the same topological node in several BATs, when the attacker used a different path to reach it (e.g. the Bayesian topological node $tn_1 \rightarrow tn_2 \rightarrow tn_4$ is different from $tn_2 \rightarrow tn_3 \rightarrow tn_4$, even if the attacker has the control of the same topological node $tn_4$ at the end.).

In the complete BAM, we thus have many Bayesian topological nodes representing the same asset of the IS.
However, what most interests a security operator is the attacks that are the most likely to compromise his assets.
Thus, as output of the consolidation of probabilities, we assign to a physical asset a probability of compromise that is the maximum of the probabilities of Bayesian topological nodes targeting the same asset.

\begin{equation*}
  P(TN_k)=\max_{i \in \{1..N\}} P_{BAT_i} (TN_k))
	= \max_{i \in \{1..N\}} (\max_{\{\forall TN_1..TN_{k-1}\}} P_{BAT_i}(TN_1,...,TN_k))
\end{equation*}

\subsection{Conditional Probability Tables}
\label{sec:cpt}

We now specify the local probability distribution associated with each node, describing the probability dependencies of a node toward his parents.
As the nodes are discrete random variables, we can describe the probability dependencies using conditional probability tables (CPT).

A Bayesian Topological node has one parent for each type of attack that can be used to compromise it.
Its probability table represents a \emph{noisy-OR}. At least one \emph{succeeded} attack step is needed to \emph{compromise} this node.
Even if no known attack step has \emph{succeeded}, there is still a little chance that an attack of this topological node may be an unknown one (e.g. a 0-day).
We denote it by $pua$.
Such a CPT is described in Appendix~\ref{sec:appendix-conditional-probability-tables} Table~\ref{cpt-topological-nodes}.

An attack step node has two types of parents:
\begin{inparaenum}[(1)]
\item one Bayesian topological node, the source of the attack, required to perform the attack;
\item one or more Bayesian condition nodes.
\end{inparaenum}
Depending on the type of attack modelled, the condition nodes may not exist for the attack node.
The $probabilityNewAttackStep$ parameter represents the fact that an attacker may have reached his objective.
Even if he has compromised the topological node and conditions are verified, it is not certain that he will attempt to propagate through the execution of a new exploit.
We describe in Appendix~\ref{sec:appendix-conditional-probability-tables} Table~\ref{cpt-attack-nodes} the CPT of a Bayesian attack node, for the exploitation of a vulnerability.

A Sensor node has only one parent, the attack node related to the sensor.
Its CPT thus contains only two values and their complement representing the $falsePositive$ and $falseNegative$ rates attached to the sensor.
The CPT of a Bayesian sensor node is described in Appendix~\ref{sec:appendix-conditional-probability-tables} Table~\ref{cpt-sensor-nodes}.

The attack source of a BAT is a Bayesian topological node without parents.
As such, it does not have a complete CPT, but only a prior probability value and its complementary.
This $attackSourceProbability$ parameter represents the \emph{a priori} probability of having an attack issued from this node.
It thus has to be set by the operators, knowing the risk that an attack starts from a topological node.
It can be deduced from a risk evaluation methodology (\eg, ISO 27005~\cite{iso2011iso}).
In a typical system, a high probability can be set to the Internet (\eg, 0.7), a medium one to servers in a demilitarised zone (internal subnetwork protected by a firewall exposing external-facing services on the Internet) (\eg, 0.4), and a small one for production database servers (\eg, 0.1).

The Attack conditions also do not have any parents.
Their probability is the probability of successful exploitation $P(c)$ associated with the condition.
It highly depends on the type of condition modelled by this node.
For example, for a condition describing the successful exploitation of at least one vulnerability of a list on a host.
The estimation of this probability of successful exploitation follows the process detailed in Section~\ref{sec:grouping-as}, with values for each vulnerability, coming from the Exploitability Metrics of the CVSS, as explained in Section~\ref{sec:tag}.

\subsection{Bayesian Attack Model usage}

We build our Bayesian Attack Model from the knowledge that the security operators have about the information system: network topology, known vulnerabilities and deployed detection sensors.
Then, we change the state of the Bayesian sensor or topological nodes according to the security events received from the sensors:

\begin{description}
  \item[Sensor Nodes:] If the sensor of an attack step exists and is deployed in the network, as long as it has not issued any alert, all related sensor nodes of the BAM (that may appear in several BATs) are set to the \emph{no alert} state.
  When the sensor raises an alert corresponding to this attack step, the Bayesian sensor nodes are set to the \emph{alert} state.
  If the sensor also gives an alert confidence probability, it is possible to set the state \emph{alert} to this probability.
  \item[Topological Nodes:] As soon as a compromise information is known for a topological node, all related Bayesian topological nodes are set to the corresponding state.
  For example, if a Host Intrusion Detection System (HIDS) says that a host is healthy, the related Bayesian topological nodes in all BATs are set to the \emph{not compromised} state.
  Conversely, if the HIDS says that a host is compromised, the related Bayesian topological nodes are set to the \emph{compromised} state.
  If the HIDS also gives a compromise probability, the \emph{compromised} state is set to this probability.
\end{description}

The Bayesian nodes for which there is no compromise information (no deployed sensor, Bayesian attack step nodes and Bayesian condition nodes) are not set in any state and their probability are updated by the Bayesian inference.

Each time the BAM changes state (when we fix nodes in a different state), we use a Bayesian network belief propagation algorithm (Lauritzen or Pearl's inference algorithm) to update the probabilities of each state at all the nodes.
Then, for each topological node of the topological attack graph, the maximum probability of the state \emph{compromised} of all related Bayesian topological nodes, provides security operators with the probability of the asset being compromised, as described in Subsection \ref{probability-consolidation}.

\subsection{Model size limitation}
\label{nbsteps-justification}

\emph{Use of a $nbSteps$ parameter to prevent performance issues:}
The main limitation when implementing this model is the combinatorial explosion of the number of nodes, due to the redundancy introduced by the cycle breaking process.
In order to improve the performance and prevent this combinatorial explosion, we limit the number of successive attack steps added to each $BAT$, according to a $nbSteps$ parameter.
Thus, we can contain the number of nodes to process in the BAM, as detailed in Section~\ref{sec:complexity-evaluation}.

\emph{Impact of the $nbSteps$ parameter on the outputs of the BAM:}
Thanks to the redundancy of the model, and as each topological node is an attack source of a $BAT$, if some attack steps are discarded in a $BAT$, they will be in another $BAT$, closer to the $BAT$ attack source.
The probabilities of Bayesian topological nodes in a $BAT$ represent the probability of the attacker exploiting this node \emph{starting from the attack source}.
As long as no attack has been detected on a path, the probability of a node compromise decreases rapidly as a function of the length of the path between the attack source and the node.
During initial probability computation, the probabilities of nodes far from the attack sources are very low.
These probabilities are below the maximum used during the probability consolidation detailed in Section~\ref{probability-consolidation} and do not have any effect on final compromise probabilities.
In that case, the $nbSteps$ parameter has no impact on final results.

The key limitation this parameter introduces is when attacks start being detected and introduced in a path.
More precisely, the limitation arises when more than two detections are injected in the model.
For example, to compute the combined impact of two detections relative to each other, they need to appear in the same BAT.
The maximum compromise probability of the topological node related to the first detection will be in the $BAT$ in which it is the attack source.
If the second detection is attached to a node that is more than $nbSteps$ away (\ie, separated with more than $nbSteps-1$ missed detections), it will not be in the same $BAT$ and these two attacks will be taken into account separately.
This will prevent the increase of probabilities of the nodes between the two detections.
Detections may be separated by other nodes without detections for two reasons: if there are not enough sensors or if there are false negatives, both undesired cases.
As a summary, the only case when the impact of the limitation of the $BAT$ depth to $nbSteps$ is significant is when there are more missed detections than $nbSteps-1$ between two successive detections for the same attack.
These assumptions are validated by the experimental validation of Section~\ref{parameter-sensitivity-analysis}.

\section{Validation}

\subsection{Complexity evaluation}
\label{sec:complexity-evaluation}

The main computation done on each Bayesian Attack Tree of the BAM is the execution of the belief propagation algorithm (probability inference), computing the probability of all nodes, according to evidences, nodes set to a specific state.
The complexity of the inference in a Bayesian network is directly linked to the number of nodes and structure of the network.
We estimate the number of nodes $M$ of a BAT, depending on $N$, the number of topological nodes in the attack graph, and $k$ the maximum number of \emph{consolidated} attack steps between two topological nodes in the topological attack graph (\ie, the maximum number of different types of attacks).
$M$ is also strongly depending on the existence of attack steps between the topological nodes.
An attack step needs the existence of at least a vulnerability and of an authorised network access, which depends totally on the monitored information system.
Thus for this complexity evaluation, we consider the worst case: there are $k$ attack steps between each pair of topological nodes.
For each attack step, we add $\approx$ 4 nodes to the BAM (sometimes few more, according to the number of conditions).
Thus, in the worst case, for each BAT, starting from an attack source, the number of nodes to add is

\begin{equation*}
M \sim 4 \times k \times (N \times \cdots \times (N-nbSteps-1)) = 4 \times k\times \frac{N!}{(N-nbSteps)!} \ = \mathcal{O}(N^{nbSteps})
\end{equation*}

The degree of the polynomial curves of the number of nodes in the BAM increases with the parameter $nbSteps$.
However, even if the number of nodes in each $BAT$ is high, the Bayesian inference can be done efficiently.
Indeed, as the structure is a \emph{polytree}, some efficient inference algorithms can be used.
For example, Pearl's belief propagation algorithm is linear in the number of nodes \cite{pearl1988probabilistic}.

Thus, for each BAT, in the worst case, the complexity of the construction and probability inference $\mathcal{C}(BAT)$ is $ \mathcal{C}(BAT) = \mathcal{O}(N^{nbSteps}) $.
Finally, for the whole BAM, as there are at most $N$ attack sources, in the worst case, the complexity of the inference in the whole model $\mathcal{C}(BAM)$ is

\begin{equation*}
\mathcal{C}(BAM) = N.\mathcal{C}(BAT) = \mathcal{O}(N^{nbSteps+1})
\end{equation*}

The calculations on each $BAT$ are independent.
So, they may be easily done in parallel, which gives in practice, $\mathcal{C}(BAM) = \mathcal{O}(N^{nbSteps})$ with $N$~processors.

\subsection{Experimental use-case-based validation}

We will first present a use-case and the scenarios that have been chosen to do the experimental validation of the BAM, then discuss the results obtained.

\subsubsection{Validation scenarios}
\label{validation-scenarios-description}

In order to validate the accuracy of the results, while keeping the scenarios simple for explanations, we implemented a real infrastructure of 11 virtual machines, for a total of a hundred vulnerabilities.
A host (that will be called host $A$, thereafter) can be attacked from the Internet, and can attack the other hosts $G$ to $J$ of its subnetwork. The latter hosts can attack hosts $A$, $C$ and $D$.
This network topology is representative of a real information system, where an ingress firewall (host $K$) protects the LAN ($E$ to $J$), and where publicly accessible servers are put in a demilitarised zone ($A$ to $D$).
The topological attack graph used to populate the BAM has been generated from a report of the vulnerability scanner Nessus, done on this infrastructure.

We apply 6 attack scenarios on this network topology, as summarised in Appendix \ref{sec:appendix-simulation-scenarios} Table~\ref{simulation-scenarios}.
The attack is carried out through three attack steps.
In the first scenario, no step is detected; it represents the basic risk of the IT system.
In scenarios 2 to 4, steps are detected and alerts are generated.
Scenarios 5 and 6 represent detection anomalies.
These scenarios represent the dynamic evolution of a system with different possible situations:
\begin{itemize}
	\item Scenarios 1, 2, 3, then 4: Normal evolution of an attack during the time.
	\item Scenarios 1, 2, then 5: Evolution of an attack in which an attack step cannot be detected (no sensor for this step).
	\item Scenarios 1, 2, then 6: Evolution of an attack in which an attack step has not been detected while there was a sensor for this step.
\end{itemize}

We assume in these scenarios that the alerts given by the sensors are binary ($alert$, $noalert$), \ie, we do not have alert confidence.

\subsubsection{Parameters default values}
\label{sec:parameter-default-values}
This use-case represents a typical critical IS.
It is managed by a security operator who often uses a vulnerability scanner.
Most vulnerabilities are known, but there is still a chance (\eg, 0.1\%) that a very motivated attacker knows a non-public vulnerability.
As the system contains known unpatched vulnerabilities, sensors are deployed to raise an alert when one of the vulnerabilities is exploited.
These sensors have a medium chance (\eg, 5\%) to raise false positives, when an attack do not succeed while being detected.
However, for the vulnerabilities for which a detection sensor is deployed, the probability of having a false negative is lower (\eg, 1\%).
The operator knows that his system is quite well protected, so it is very unlikely that an attack occurs with more than 2 undetected steps (\eg, $nbSteps$ can be set to 3).
Most attacks may come from the Internet (\eg, probability of the Internet being a source of attack of 70\%), even if internal hosts may also be a new source of attacks (undetected phishing, malicious employee, \etc) with a lower probability (\eg, 10\%)).
Finally, as valuable machines are not deeply protected (they can be reached in 3 steps from the Internet), the probability that the attacker propagates through a new attack step is medium (\eg, 30\%).
After an attack, he may have already found what he was looking for.
Default values of the parameters used for this use case are summarised in Appendix~\ref{sec:appendix-default-values-parameters} Table~\ref{default-parameters-values}.

\subsubsection{Results and analysis}

The Bayesian Attack Model was implemented in Java, using the SMILE Bayesian Network library~\cite{druzdzel1999smile}.
The results of the compromise probabilities of each topological node calculated by the BAM, for each scenario, are shown in Appendix~\ref{sec:appendix-validation-results} Figure~\ref{results-scenarios}.
The first scenario is the basic risk.
The only host that has a \emph{medium} risk is the Internet.
The other hosts have a \emph{not-significant} risk.
In the scenarios 2, 3 and 4, the sensors corresponding to the 3 steps attack are set progressively.
Each new sensor set as \emph{detected} confirms the attack that is currently happening and increases the compromise probability of the previous and future states.
For example, in scenario 4, the Internet, and the 3 victim hosts are in the \emph{high}-risk zone.
In scenario 5, and scenario 6, when there is a missing detection or a false negative / false positive, the probabilities of an ongoing attack are lower, but higher than the basic risk, and the probabilities of scenario 2, that should precede this state. So, a security operator may investigate the appropriate machines to confirm or disprove the attack.

\subsubsection{Parameter sensitivity analysis}
\label{parameter-sensitivity-analysis}

Several parameters can be customised in the BAM (\cf\ Section \ref{sec:parameter-default-values}).
We summarise in Appendix~\ref{sec:appendix-sensitivity-analysis}, Table~\ref{sensitivity-analysis} the results of the sensitivity analysis of these parameters with the range of variation that we find appropriate for the given parameters (range of values that may occur in real-life).
The false positives and negatives rates vary from 0 to 30\%, because beyond, their values are meaning-less (\eg, a vulnerability signature with more 30\% false positive is useless).
The number of successive steps varies from 1 (its minimum) to 4 (the maximum possible number of successive attack steps for this use-case).
Sources probabilities vary from 0 to 1, as, according to the context, all values may be possible.
The probability of having an exploitation of an unknown vulnerability is low (15\% is a far upper bound).
The probability of the attacker making a new attack step is difficult to estimate.
We thus need to study the impact of this parameter on its whole possible variation interval (0 to 100\%).

The most interesting result of this analysis is the \emph{ranking influence} describing the impact of the variation of a parameter on the rank of topological nodes probabilities (on the whole parameter variation range, for the 6 scenarios).
This rank will determine the priorities of security operators in their IS.
The \emph{probability influence} describes the effect of the variation of the parameters on the absolute value of the topological nodes probability.
The only parameter that has an impact on the rank of the topological nodes probabilities is $probabilityOtherHosts$.
However, this parameter can be estimated quite accurately with a risk analysis methodology, which gives the security risk of each topological node, according to its position in the information system. All other parameters do not have any effect on the ranking on their whole variation range, which is a comforting result.
Four parameters have a medium impact on the absolute value of the compromise probabilities of topological nodes.
With a medium uncertainty on such parameters (\eg, 0.2), the variation of the absolute value of the probabilities is medium (\eg, up to 0.2).
Other parameters have a low impact on absolute values of probabilities.
With a medium uncertainty on such parameters (\eg, 0.2), the variation of the absolute value of the probabilities is low (\eg, up to 0.02).
So, absolute value of probabilities may be a little impacted by uncertainty on parameters, but rank is mostly not impacted by the variation of the parameters.

\subsection{Performance evaluation}

\label{sec:performance}

In order to dynamically assess the risk of a system, the BAM has to be evaluated each time a correlated alert, or a set of correlated alerts is received: the sensors and topological nodes are set in their new states, then the probabilities are updated.
The duration of such a process needs to be quite fast (around 1 minute is good), for the operator to properly understand the risk in operational time.
We simulate random network topologies with different parameters (number of hosts, subnets, vulnerabilities and network services and connectivity between subnets) to evaluate the performance of the BAM.
We generate the TAGs related to the topologies.
Then, we generate random attack scenarios with seven successive attack steps. 
Finally, we evaluate the BAM on the different scenarios.

\begin{figure}[!h]
  	\centering
	\includegraphics[width=8cm]{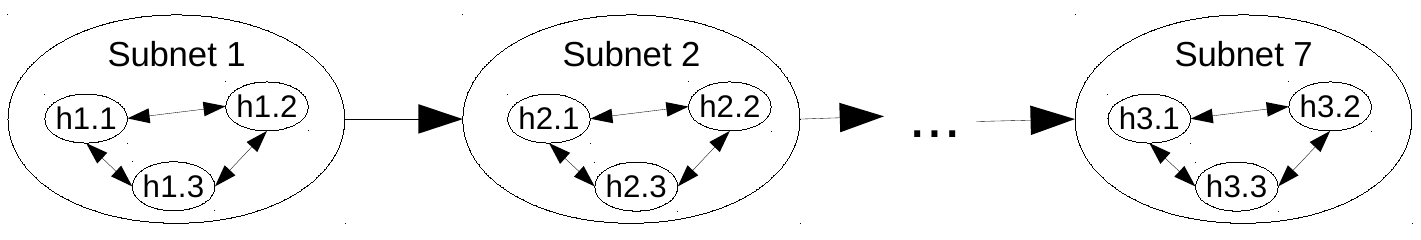}
	\caption{\label{simulation-topology} Network topology for simulations}
\end{figure}
We generate random topologies, as depicted in Figure~\ref{simulation-topology}, containing from 1 to 70 hosts, in 7 subnets.
These topologies are representative of a real network in which defense in depth is implemented: all the hosts of a subnet have access to all the hosts of a deeper subnet.
In each subnet, all accesses between hosts are authorized.
Each host has 30 random vulnerabilities for a maximum total of around 2000 vulnerabilities.
The results of the duration in seconds of the BAM generation and the inference after the evaluation of one scenario of 7 successive attack steps, on these topologies, is displayed in Appendix~\ref{sec:appendix-performance-evaluation-results} Figure~\ref{simulation-durations}.
The parameters of the BAM are in the default values detailed in Section~\ref{default-parameters-values}.
This simulation shows that for medium-sized topologies (up to 70 hosts) the duration of the Bayesian Attack Model generation and of the inference remains acceptable ($<$ 1 minute 30 seconds) on a laptop-class computer.

Even if the number of topological nodes of these simulations is limited (70 hosts), it could be extended to much bigger IS, by clustering together identical templates of servers or of client machines in one topological node, as they possess the same vulnerabilities and authorised accesses and thus behave in a similar way in the BAM.
Even with 60 assets in the topological attack graph with, for example, 10 templates of client machines, 30 of network servers, and 20 of business application servers, it is possible to model a usual big-sized IS.

\subsection{Accuracy evaluation}

To evaluate the accuracy of the results (\ie, how close the probabilities are to the truth), we simulate attack scenarios on the random topologies presented in Section~\ref{sec:performance} and compare the theoretical results with the outputs of the BAM.
The results are shown in Appendix~\ref{sec:appendix-accuracy-evaluation-results} Figure~\ref{simulation-accuracy}.
We compare the theoretical results known in the scenarios with the results of the BAM.
In each scenario, we know the nodes that are compromised and healthy, \ie, nodes with a theoretical probability of respectively 1 and 0.
Then, we assess if the BAM probabilities of compromised nodes are close to 1, and if the BAM probabilities of healthy nodes are close to 0.
The plot shows the maximum errors (in terms of distance to the theoretical values 1 and 0) of compromised and healthy nodes.
This figure shows a large free space between the errors on compromised hosts and the errors on healthy hosts. This means that if there are no false-positives nor false-negatives in the detection inputs of the BAM, it allows to distinguish exactly healthy and compromised hosts, for example with a boundary at the probability of 0.5.
So, there are no false negatives nor false positives introduced by the BAM.
The graphical difference of the results between the values for a low number of hosts and high number of hosts is probably due to the random attack scenarios that may be shorter when there are not enough hosts.

\section{Related Work}

Many people proposed enhancements to improve attack graphs with Bayesian networks, to use them for dynamic risk assessment~\cite{Qin:2004fi,Liu:2005un,Xie:2010ks}.
However, they do not describe how they manage cycles that are inherent to attack graphs.
In~\cite{Xie:2010ks}, Xie~\etal present an extension of MulVAL attack graphs using Bayesian networks, but they do not mention how to manage the cycle problem, while MulVAL attack graphs frequently contain cycles.
In the same way, in~\cite{Frigault:2008jp}, Frigault and Wang do not mention how they deal with the cycle problem constructing Bayesian attack graphs.
In~\cite{Liu:2005un}, Liu and Man assert that to delete cycles, they assume that an attacker will never backtrack.
The same assumption is used by Poolsappasit~\etal in~\cite{poolsappasit2012dynamic}.
However, they both do not present how they deal with this assumption to keep all possible paths in the graph, while deleting cycles.
We propose here a novel model that explodes cycles in the building process, keeping all possible paths while deleting the cycles, to compute the Bayesian inference.

The Bayesian model presented by Xie~\etal in~\cite{Xie:2010ks} is based on logical attack graphs.
It is thus very verbose and can be huge for real information systems.
In~\cite{Liu:2005un}, Liu and Man's model is a topological graph, in which are added violation states.
It is thus quite compact, but does not detail the attacks, their conditions and, mainly, the sensors that can change state.
Thus, the only observations that can be set on this model are observations on topological nodes.
The model we present is a topological model.
So, it is much more compact than those based on logical attack graphs.
However, it contains the logical conditions necessary to carry out the attacks, in order to keep all information important to model attacks, and add sensor nodes that can be activated with detections.
Moreover, we also add several improvements (attack nodes gathering, polytree structure of BAT, etc.) that either reduce the size of the graph structure or improve the performance of the inference.
We thus constrain the size of the graph in which we do Bayesian inference, while conserving all paths by linearising cycles.

The experimental validation we did on the Bayesian Attack Model is on a real topology of a complexity similar or superior to what was done in the literature and on simulated topologies that are far bigger than the state of the art.
For example, Xie~\etal assess their model on a topology of 3 hosts and 3 vulnerabilities~\cite{Xie:2010ks},
Liu and Man on a topology of 4 hosts and 8 vulnerabilities~\cite{Liu:2005un}.
The real world examples used by Frigault and Wang in~\cite{Frigault:2008jp} contain at most 8 vulnerabilities on 4 hosts.
The test network used by Poolsappasit~\etal in~\cite{poolsappasit2012dynamic} contains 8 hosts in 2 subnets, but with only 13 vulnerabilities.
Thanks to our polytree model, we successfully run our Bayesian Attack Model efficiently on simulated topologies with up to 70~hosts for a total of more than 2000 vulnerabilities.

\section{Conclusion and Future Work}

We present in this paper a new Bayesian Attack Model (BAM), representing all the possible attacks in an information system.
This model enables dynamic risk assessment.
It is built from a topological attack graph, using already available information.
Sensor nodes can be activated by dynamic security events to update the compromise probabilities of topological assets, which rank the risk level of ongoing attacks.
This model handles the cycles that are inherent to attack graphs and thus is applicable to any information system, with multiple potential attack sources.
The cycle breaking process significantly increases the number of nodes in the model, but thanks to the polytree structure of the Bayesian networks we build, the inference remains efficient, for medium information systems.
In order to be able to use the Bayesian Attack Model for bigger information systems, future work will investigate how the usage of a hierarchical topological attack graph can be appropriate to build the Bayesian Attack Model.

\bibliographystyle{splncs03}
\bibliography{references}

\appendix
\clearpage

\section{Appendix: Detail of a Bayesian attack step}
\label{sec:appendix-detail-attack-step}

\begin{figure}[h!]
	\caption{\label{bayesian-attack-step} Bayesian attack step}
	\begin{center}
		\includegraphics[width=8cm]{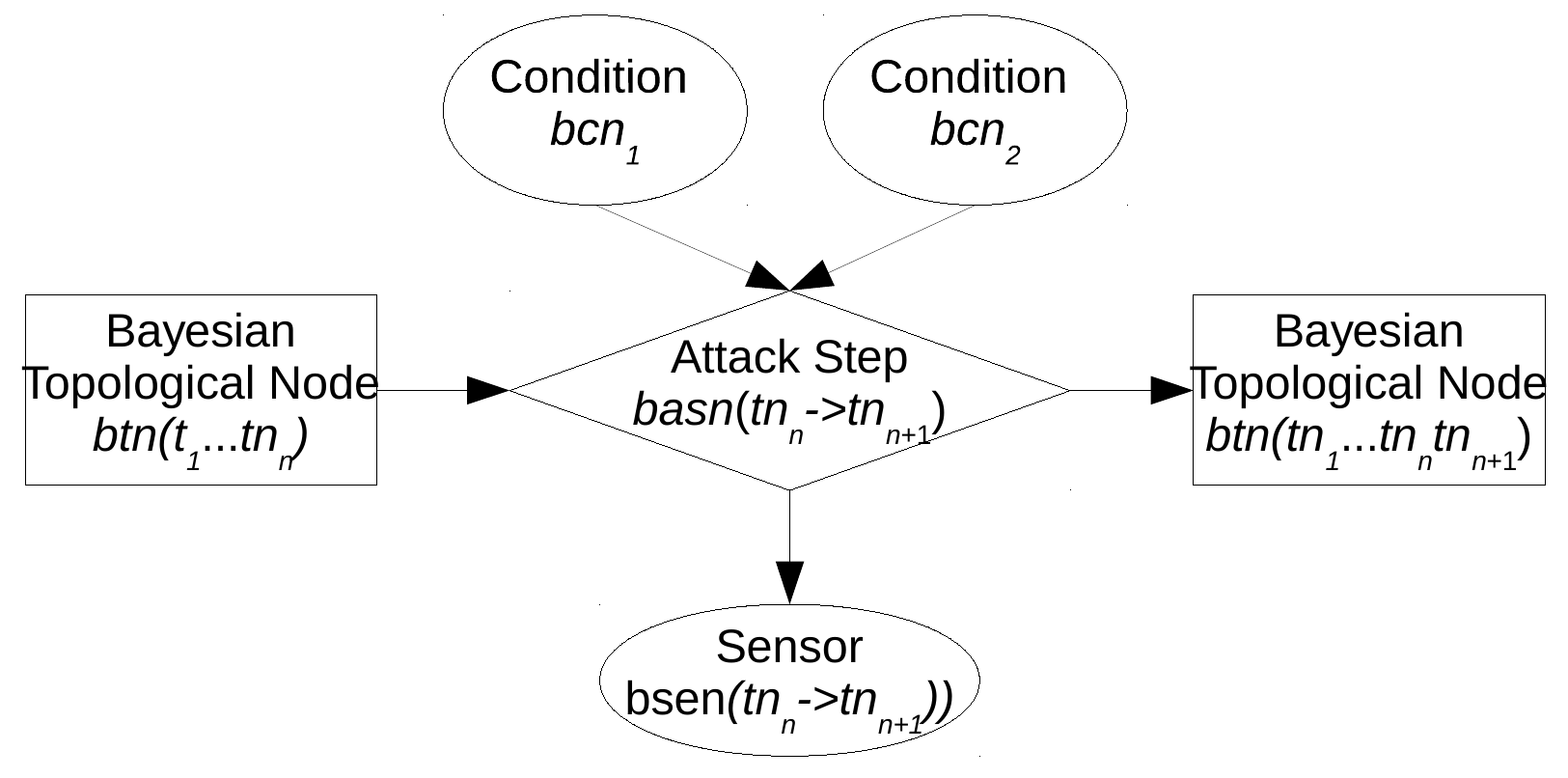}
	\end{center}
\end{figure}

Figure~\ref{bayesian-attack-step} shows the details of the representation of an attack step from $tn_n$ (source) to $tn_{n+1}$ (target).
It is composed of a Bayesian attack step node that binds a Bayesian topological node to another one.
This Bayesian attack step has two conditions ($bcn_1$ and $bcn_2$) and a sensor ($bsen$).

\section{Appendix: Conditional Probability Tables}
\label{sec:appendix-conditional-probability-tables}

In this appendix, we detail the conditional probability tables (CPTs) associated with the nodes of the Bayesian Attack Model.
Each node with at least one parent is associated with a CPT which depends on its type of node.
In these tables, the first lines represent all possible states of the parents.
The last lines contain the probabilities of each state of the child node according to the states of its parents.

Table~\ref{cpt-topological-nodes} shows the CPT of a Bayesian topological node, according to the states of its parents: Bayesian attack step nodes.
It represents a noisy OR: an OR with a small residual probability (the $probabilityUnknownAttack$ parameter).

\begin{table}[h!]
  \caption{CPT of a Bayesian topological node\label{cpt-topological-nodes}}
  \centering
  \begin{tabular}{|p{4.6cm}||p{1.5cm}|p{1.5cm}|p{1.5cm}|p{1.5cm}|}
		  \hline
		  $\textbf{Bayesian attack step node}$ \textbf{1} & \emph{Succeeded} & \emph{Failed} & \emph{Succeeded} & \emph{Failed} \\ \hline
		  $\textbf{Bayesian attack step node}$ \textbf{2} & \multicolumn{2}{c|}{\emph{Succeeded}} & \multicolumn{2}{c|}{\emph{Failed}} \\ \hline \hline
		  $\textbf{Bayesian topological node}$ & \multicolumn{4}{c|}{}\\ \hline
		  \emph{Compromised} & 1 & 1 & 1 & $pua$ \\ \hline
		  \emph{NotCompromised} & 0 & 0 & 0 & $1 - pua$ \\ \hline
   \end{tabular}

	\textbf{with} $pua = probabilityUnknownAttack$.

\end{table}

Table~\ref{cpt-attack-nodes} shows the CPT of a Bayesian attack step node, for the exploitation of a vulnerability, according to the states of its parents: a Bayesian topological node and a Bayesian condition node.
It represents an AND on the parents with the $probabilityNewAttackStep$ parameter, when all conditions are fulfilled.

\begin{table}[h!]
  \caption{CPT of a Bayesian attack step node "exploitation of a vulnerability"}\label{cpt-attack-nodes}
  \centering
\begin{tabular}{|p{4.6cm}||p{1.5cm}|p{1.5cm}|p{1.5cm}|p{1.5cm}|}
		  \hline
		  $\textbf{Bayesian topological node}$ & \emph{Comp} & \emph{NotComp} & \emph{Comp} & \emph{NotComp} \\ \hline
		  $\textbf{Bayesian condition node}$ & \multicolumn{2}{c|}{\emph{Succeeded}} & \multicolumn{2}{c|}{\emph{Failed}} \\ \hline \hline
		  $\textbf{Bayesian attack step node}$ & \multicolumn{4}{c|}{}\\ \hline
		  \emph{Succeeded} & $pnas$ & 0 & 0 & 0 \\ \hline
		  \emph{Failed} & $1-pnas$ & 1 & 1 & 1 \\ \hline
   \end{tabular}

	 \textbf{with} $Comp$ = State $Compromised$; $NotComp$ = State $NotCompromised$; \\ $pnas$ = $probabilityNewAttackStep$.
\end{table}

Finally, Table~\ref{cpt-sensor-nodes} shows the CPT of a Bayesian sensor node, according to the state of its parent: a Bayesian attack step node.
It represents the potential false positive and false negative rates of the sensor.

\begin{table}[h!]
  \caption{CPT of a Bayesian sensor node}\label{cpt-sensor-nodes}
  \centering
\begin{tabular}{|p{4.6cm}||p{3cm}|p{3cm}|}
		  \hline
		  $\textbf{Bayesian attack step node}$ & \emph{Succeeded} & \emph{Failed}\\\hline \hline
		  $\textbf{Bayesian sensor node}$ & \multicolumn{2}{c|}{}\\ \hline
		  \emph{Alert} & $1 - falseNegative$ & $falsePositive$ \\ \hline
		  \emph{NoAlert} & $falseNegative$ & $1 - falsePositive$\\\hline
   \end{tabular}
\end{table}

\section{Appendix: Simulation scenarios}
\label{sec:appendix-simulation-scenarios}

Table~\ref{simulation-scenarios} shows the detection scenarios applied on the use-case.
In the first scenario, no step is detected; it represents the basic risk of the IT system.
In scenarios 2 to 4, steps $I \rightarrow A$, $A \rightarrow G$ and $G \rightarrow D$ are progressively detected and alerts are generated.
Scenarios 5 and 6 represent detection anomalies on $A \rightarrow G$ (no sensor information or false negative).

\begin{table}[!h]
  \caption{\label{simulation-scenarios} Simulation scenarios}
  \centering
  \begin{tabular}{|c||c|c|c|p{8cm}|}
    \hline
    \textbf{Scenario}	& $I \rightarrow A$ 	& $A \rightarrow G$ 	& $G \rightarrow D$ 	& \textbf{Comment} \\
    \hline \hline
    \textbf{1} 		& $\times$			& $\times$			& $\times$			& Basic risk \\ \hline
    \textbf{2} 		& $\checkmark$		& $\times$			& $\times$			& First alert \\ \hline
    \textbf{3} 		& $\checkmark$		& $\checkmark$		& $\times$			& Second alert \\ \hline
    \textbf{4} 		& $\checkmark$		& $\checkmark$		& $\checkmark$		& Third alert \\ \hline
    \textbf{5} 		& $\checkmark$		& $O$				& $\checkmark$		& No information available for the second step (=no sensor) \\ \hline
    \textbf{6} 		& $\checkmark$		& $\times$			& $\checkmark$		& No detection for the second step \\ \hline
  \end{tabular}

\textbf{Caption} $I \rightarrow A$: Attack from the Internet to host $A$; $A \rightarrow G$: Attack from host $A$ to host $G$; $G \rightarrow D$: Attack from host $G$ to host $D$; $O$: No values set (=no sensor); $\checkmark$:~Sensor node set to $alert$; $\times$:~Sensor node has been set to $no alert$.
\end{table}

\newpage

\section{Appendix: Default values of the parameters}
\label{sec:appendix-default-values-parameters}

Table~\ref{default-parameters-values} presents all the parameters of the Bayesian Attack Model: \textit{probabilityUnknownAttack}, \textit{falsePositive}, \textit{falseNegative}, \textit{nbSteps}, \textit{probabilityInternet}, \textit{probabilityOtherHosts}, and \textit{probabilityNewAttackStep}.
Each parameter is associated with its description and the default value that was chosen for the use-cases.

\begin{table}[h!]
  \caption{\label{default-parameters-values} Default values of the parameters used in the BAM}
  \centering
  \begin{tabular}{|p{2.5cm}||p{1.4cm}|p{3.1cm}|p{5.5cm}|}
    \hline
    \textbf{Parameter name} & \textbf{Default value} & \textbf{Meanings} & \textbf{Default value explanation}\\
    \hline \hline
    \textit{probability-UnknownAttack} & $0.001$ & Probability that an unknown attack occurs. & Very small probability of having a 0-day, a unknown vulnerability. \\ \hline
    \textit{falsePositive} & $0.05$ & False positive rate of each sensor. & Sensors may raise an alert, even if the attack has not succeeded.\\ \hline
		\textit{falseNegative} & $0.01$ & False negative rate of each sensor. & This value is smaller as it only concerns vulnerabilities for which a sensor has been deployed.\\ \hline
    \textit{nbSteps} & $3$ & Number of successive attack steps to keep. & Allow to recognise multi-step attacks with at most 2 missing alerts. \emph{C.f.}~Subsection~\ref{nbsteps-justification} for full explanation.\\ \hline
    \textit{probability-Internet} & $0.7$ & A priori probability of an attack coming from the Internet. & The internet is the main source of attacks. Thus, 70\% of chances of being a source of attack, 30\% not to be a source.\\ \hline
    \textit{probability-OtherHosts} & $0.1$ & A priori probability of an attack issued from an internal host. & An internal host may issue an attack. Thus, 10\% of chances of being a source of attack, 90\% not to be a source.\\ \hline
    \textit{probability-NewAttackStep} & $0.3$ & Probability that the attack propagates through a new attack step. & 70\% of chance that the attacker does not continue his attack. He may have already found what he was looking for.\\ \hline
  \end{tabular}
\end{table}

\section{Appendix: Validation results}
\label{sec:appendix-validation-results}

\begin{figure}[h!]
  	\centering
	\includegraphics[width=12cm]{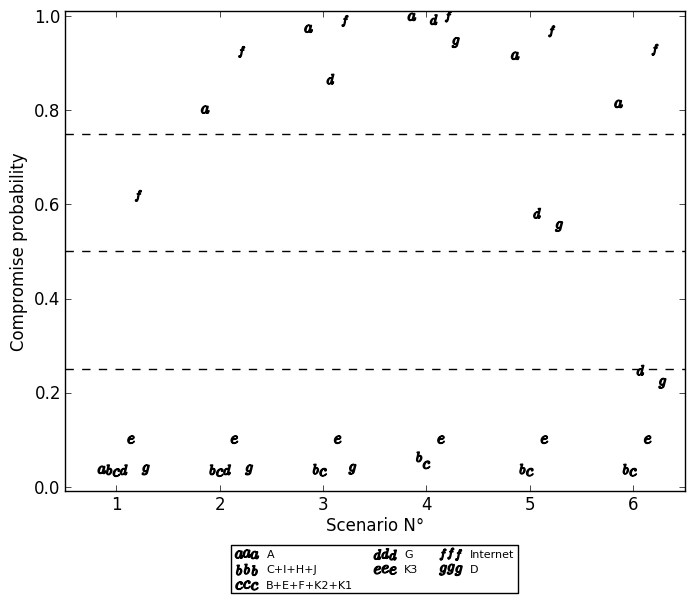}
	\caption{\label{results-scenarios} Results for each scenario}
	\small{For readability of the results, the hosts having the same value (more or less $10^{-10}$) for all the scenarios have been grouped on one point, and the points are spread around the scenario number.}
\end{figure}

Figure~\ref{results-scenarios} shows the results of the BAM for the six scenarios of the use case presented in Subsection~\ref{validation-scenarios-description}. Markers represent hosts of the topology, and the ordinate is their compromise probability in the abscissa scenario. The horizontal lines give some idea of the threshold that could be taken to define the compromise risk level of the hosts. For example, the hosts under the lowest line ($probability \leq 0.25$) have a \emph{not-significant} risk of being compromised, above the lowest line ($0.25 < probability \leq 0.50$) have a \emph{low} risk, above the second line ($0.50 < probability \leq 0.75$) have a \emph{medium} risk, and above the upper line ($0.75 < probability$) have a \emph{high} risk of being compromised.

\newpage
\section{Appendix: Parameters sensitivity analysis results}
\label{sec:appendix-sensitivity-analysis}

Table~\ref{default-parameters-values} summarises the sensitivity analysis of the parameters of the Bayesian Attack Model.
We give in this table the range of variation that we find appropriate for the parameters.
Then, we study the influence of each parameter in its whole variation range on the ranking between the compromise probability of the hosts, and on the value of the probabilities.

\begin{table}[h!]
	\begin{small}
  \caption{\label{sensitivity-analysis} Sensitivity analysis of the parameters of the BAM}
  \centering
  \begin{tabular}{|p{1.9cm}||p{1.5cm}|p{2cm}|p{2cm}|p{5cm}|}
    \hline
    \textbf{Name} & \textbf{Variation range} & \textbf{Ranking influence} & \textbf{Proba influence} & \textbf{Comment}\\
    \hline \hline
    falseNegative & $[0.0-0.3]$& No impact & Almost no impact & \\ \hline
    falsePositive & $[0.0-0.3]$& No impact & Medium impact in scenarios 3, 4 and 5 (decrease) & Impact for low values (between 0 and 0.05). Effect amplified by the number of detections set.\\ \hline
    nbSteps & $[[1-4]]$& No impact & Medium impact in scenarios 4, 5 and 6& Parameter sensitive particularly for scenarios that contain attacks longer than $nbSteps$.\\ \hline
    probability
    Internet & $[0.0-1.0]$& No impact (except the Internet)& Little impact (increase) & Probability of hosts attacked directly from the Internet increases with the increase of this parameter.\\ \hline
    probability
    OtherHosts & $[0.0-1.0]$& Medium Impact (different probability growth curve)& Medium impact (increase) & Probability of all hosts (except the Internet) increase with the increase of such parameter. Stronger increase for hosts attackable from many hosts.\\ \hline
    probability
    NewAttack & $[0.0-0.15]$& No impact (except the Internet) & Low impact (increase) & With the increase of this parameter, the probability of attackable hosts increases slowly, as it is more probable that they are attacked, using attacks that are not known and cannot be detected. \\ \hline
    probability
    NewAttackStep & $[0.0-1.0]$& No impact & Medium impact (increase then decrease, maximum around 0.3) & When this parameter is small (increase from 0 to 0.3), the parameter represents that even if an attack is possible, it may not happen. Then (decrease from 0.3 to 1) it represents that even if an attacker has compromised a host, he may not do another attack.\\ \hline
  \end{tabular}
\end{small}
\end{table}

\newpage
\section{Appendix: Performance and evaluation results}
\label{sec:appendix-performance-evaluation-results}

\begin{figure}[h!]
  \caption{\label{simulation-durations} Duration in seconds, according to the number of hosts}
	\includegraphics[height=5cm,width=7cm]{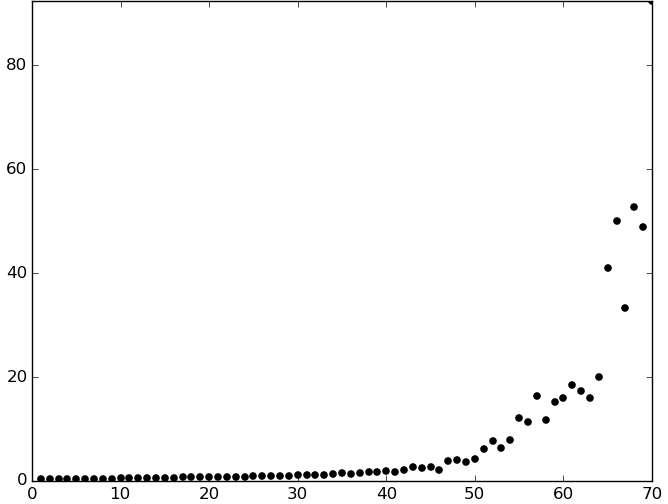}
\end{figure}

Figure~\ref{simulation-durations} presents the results of the duration in seconds of the Bayesian Attack Model generation and the inference after the evaluation of one scenario of 7 successive attack steps, on random simulated topologies from 1 to 70 hosts.

\label{sec:appendix-accuracy-evaluation-results}

\begin{figure}[!h]
  	\centering
	\includegraphics[height=5.5cm,width=7cm]{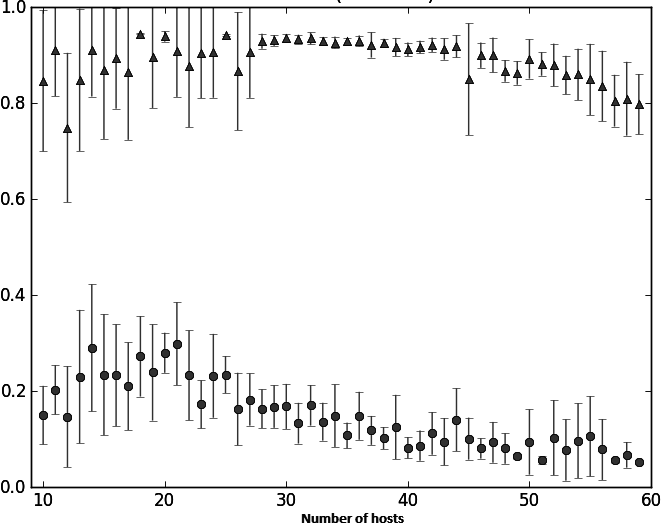}
	\caption{\label{simulation-accuracy} Accuracy of the results of the BAM according to the number of hosts}
\end{figure}

Figure~\ref{simulation-accuracy} presents the results of an accuracy evaluation of the BAM on random simulated topologies.
The curve with triangles represent the mean and standard deviation, during 10 simulations, of the minimum probability of the hosts known as compromised.
The curve with circles represent the mean and standard deviation, during 10 simulations, of the maximum probability of the hosts known as healthy.
In other words, this graph shows the maximum errors (in terms of distance to the theoretical values 1 and 0) of compromised and healthy nodes.

\end{document}